\documentclass[reprint,amsmath,amssymb,aps,prx]{revtex4-2}

\usepackage{graphicx}
\usepackage{dcolumn}
\usepackage{bm}
\usepackage{booktabs}
\usepackage{array}
\usepackage{paralist}
\usepackage{verbatim}
\usepackage{float}
\usepackage{booktabs}
\usepackage{color,soul}

\usepackage{fancyhdr}
\begin{document}

\title{High-Temporal-Resolution Measurements of the Impacts of Ionizing Radiation on Superconducting Qubits}
\author{Jihee Yang}
\affiliation{Northrop Grumman, Linthicum, Maryland 21090, USA}
\author{Thomas J. Carroll}
\affiliation{Northrop Grumman, Linthicum, Maryland 21090, USA}
\author{Philip Mason}
\affiliation{Northrop Grumman, Linthicum, Maryland 21090, USA}
\author{Robert Schwartz}
\affiliation{Northrop Grumman, Linthicum, Maryland 21090, USA}
\author{Kenneth M. O'Hara}
\affiliation{Northrop Grumman, Linthicum, Maryland 21090, USA}
\author{Jennifer Lund}
\affiliation{Northrop Grumman, Linthicum, Maryland 21090, USA}
\author{Michael Gottschalk}
\affiliation{Northrop Grumman, Linthicum, Maryland 21090, USA}
\author{Timothy Stephenson}
\affiliation{Northrop Grumman, Linthicum, Maryland 21090, USA}
\author{Lawrence H. Friedman}
\affiliation{Northrop Grumman, Linthicum, Maryland 21090, USA}
\author{Francisco Yumiceva}
\affiliation{Northrop Grumman, Linthicum, Maryland 21090, USA}
\author{Justin Hackley}
\affiliation{Northrop Grumman, Linthicum, Maryland 21090, USA}
\author{Aurelius L. Graninger}
\affiliation{Northrop Grumman, Linthicum, Maryland 21090, USA}
\author{Chris Rotella}
\affiliation{Northrop Grumman, Linthicum, Maryland 21090, USA}
\author{Pat Warner}
\affiliation{Northrop Grumman, Linthicum, Maryland 21090, USA}
\author{Jonathan M. Cochran}
\affiliation{Northrop Grumman, Linthicum, Maryland 21090, USA}
\author{Adam V. Bruce}
\affiliation{Northrop Grumman, Linthicum, Maryland 21090, USA}
\author{Melody Wagner}
\affiliation{Northrop Grumman, Linthicum, Maryland 21090, USA}
\author{James Wenner}
\affiliation{Northrop Grumman, Linthicum, Maryland 21090, USA}
\author{Stan Steers}
\affiliation{Northrop Grumman, Linthicum, Maryland 21090, USA}
\author{Christopher Moore}
\affiliation{Northrop Grumman, Linthicum, Maryland 21090, USA}
\author{Alex Marakov}
\affiliation{Northrop Grumman, Linthicum, Maryland 21090, USA}
\author{Bradley G. Christensen}
\affiliation{Northrop Grumman, Linthicum, Maryland 21090, USA}
\date{\today}

\begin{abstract}
We measure the effect of ionizing radiation on superconducting qubits with a timing resolution of 1~$\mu$s using microwave kinetic inductance detectors (MKIDs) fabricated on the same substrate. We observe no correlation between two-level system (TLS) scrambling events and ionizing radiation events detected with the MKIDs, suggesting TLS scrambling events may not arise from ionizing radiation and instead the previously reported apparent correlation may be due to events without sufficient energy to trigger our MKIDs. We characterize the fast-time system recovery of transmons following a radiation event, where we observe the recovery of the enhanced qubit relaxation and excitation to be well-described by an exponential recovery to the baseline quasiparticle density, with a characteristic time of $13\pm1\ \mu$s, and a peak quasiparticle density at the junction per deposited energy of 240\,$/\mu m^3/$MeV. The fast recovery is consistent with literature reported values for Nb-based devices with direct injection of 2$\Delta_{\text{Al}}$ phonons, demonstrating the recovery is strongly dependent on the proximity of niobium to the junction.
\end{abstract}

\maketitle

\section{\label{sec:intro} Introduction}

The effect of ionizing radiation on superconducting qubits has garnered interest due to the  potentially catastrophic impact on quantum computing \cite{McEwen_2021,Thorbeck_2023, mcewen2024resistinghighenergyimpactevents, Wilen_2021, martinis2021savingsuperconductingquantumprocessors,Loer_2024} and challenges posed for quantum sensing \cite{Karatsu_2019} platforms. When ionizing radiation traverses through a semiconducting substrate (e.g., silicon, sapphire), the radiation deposits an average energy of approximately 1 MeV per 1 mm of substrate. The energy is initially in the form of e-h pairs that quickly downconvert and recombine to cause a burst of phonons in the substrate. The phonons will downconvert to around 5~meV \cite{martinis2021savingsuperconductingquantumprocessors}, well above the superconducting gap of materials used in qubit devices (e.g., aluminum, tantalum, niobium). These phonons can thus break Cooper pairs and generate quasiparticles (QPs) in superconducting films, which can couple to the qubit photon to tunnel across the junction, leading to relaxation or excitation of the qubit.

In addition to the direct impact to qubit device performance, Thorbeck et al. showed two-level systems (TLS) \cite{Muller_2019} present in amorphous interfaces may also have dynamics driven by ionizing radiation events \cite{Thorbeck_2023}. TLS can have frequencies near the qubit frequency with coupling strengths up to 100 MHz, depending on the proximity of the defect to high-field regions of the qubit, e.g., the junction \cite{Martinis_2005}, and generally provide the dominant loss mechanism in superconducting qubits. The microscopic origin of these defects is still not understood; the dynamics of defects in the presence of a phonon burst provide additional insight into their physical nature.

Various detection methodologies have been used to characterize the impact and dynamics of ionizing radiation on superconducting qubits: charge-jumps in charge-sensitive transmons \cite{Thorbeck_2023, Wilen_2021}, rapid charge-parity changes in charge-sensitive transmons \cite{Diamond_2022}, and correlated qubit decays in large-scale arrays \cite{McEwen_2021,harrington2024synchronousdetectioncosmicrays}. However, these platforms rely on qubit measurements, which increase the complexity of the test platforms. Furthermore, these platforms are fundamentally sensitive to events at energy scales well below that of ionizing radiation; therefore, while the majority of detected events are likely due to ionizing radiation, other cryogenic effects may obfuscate interpretations of the dynamics caused by ionizing radiation.

\begin{figure}[h!]
	\includegraphics[width=0.5\textwidth]{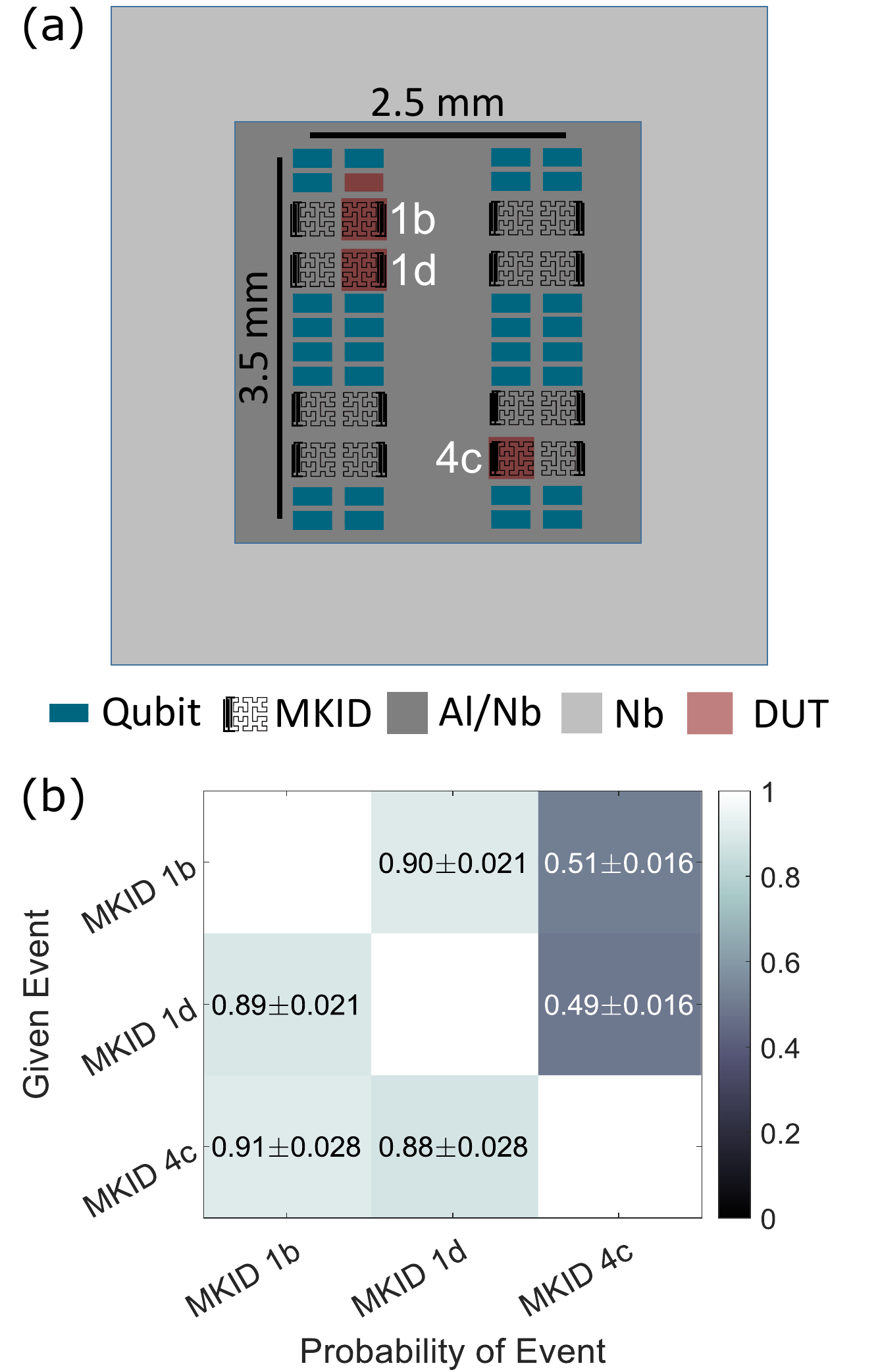}
	\caption{\label{fig:Figure1} (a) Cartoon of the device. The chip is broken into quadrants, with each quadrant having eight qubits surrounding four MKIDs. The red regions indicated devices used to collect data shown in this manuscript. (b) Conditional probabilities between three MKIDs as labeled in (a). The center of MKID 4c is 2.3 mm away from MKID 1d. The conditional matrix is consistent with ionizing radiation events being chip wide, with MKID detection efficiencies of 0.90, 0.89, and 0.50 for MKID 1b, 1d, and 4c, respectively.}
\end{figure}

Here, we present a test platform with microsecond temporal resolution achieved by incorporating microwave kinetic inductance detectors (MKIDs) \cite{Day2003-ga} on the same chip as qubits (Fig.~\ref{fig:Figure1}), which only requires a single additional rf feedline to operate. We demonstrate the capability of the test platform by monitoring transmons for enhanced relaxation, excess excitations, and TLS dynamics relative to ionizing radiation events. We observe a degradation of $T_1$ with a $13\pm1\ \mu$s characteristic recovery time, and dynamics consistent with a trapping (exponential recovery) model, similar to Ref.~\cite{Yelton_2024}. We additionally observe qubit excitations with a similar timescale for recovery, which occur due to the excess energy of the QPs that diffuse from the niobium (Nb) films, used for the qubit capacitors, onto the aluminum (Al) junction leads, and subsequently reach the Al/AlOx/Al junction itself prior to the QPs sufficiently downconverting to the Al band edge.

Finally, we observe TLS dynamics uncorrelated with ionizing radiation across 371 detected TLS scrambling events. Our MKIDs detection threshold is around 40 keV of energy deposited on the chip, which results in 25~eV, or $7.5\times 10^4$ QPs, in the superconducting film. This energy scale can likely only be achieved with ionizing radiation. Previous reports correlating ionizing radiation with TLS scrambling events \cite{Thorbeck_2023,bratrud2024measurementcorrelatedchargenoise} used charge jumps in charge-sensitive transmons to identify an ionizing radiation event. These devices rely on a permanent change to the local charge environment, which may be caused by lower-energy physical processes in addition to ionizing radiation. Multiple groups have observed oddities with charge-jumps and quasiparticle bursts. Novel charge jump behavior has been observed in tantalum-based devices \cite{tennant2021lowfrequencycorrelatedcharge}, charge jump rates have been observed to be a factor of 7 times higher than expected in an underground facility \cite{bratrud2024measurementcorrelatedchargenoise}, and quasiparticle bursts have been observed to be time dependent \cite{bratrud2024measurementcorrelatedchargenoise}. While local trapping of free-charge caused by ionizing radiation is one method of causing a charge jump, strain relaxation, e.g., could alter the qubit charge environment through rearranging TLS or through freeing a locally trapped charge, both of which would register as a charge jump. The observed phenomena, together with our results demonstrating no correlation between TLS scrambling and ionizing radiation, suggest that a different physical process is causing the charge jumps that correlate with TLS dynamics.

\section{\label{sec:device}Device Design \& Operation}

\begin{figure}[h]
	\includegraphics[width=0.5\textwidth]{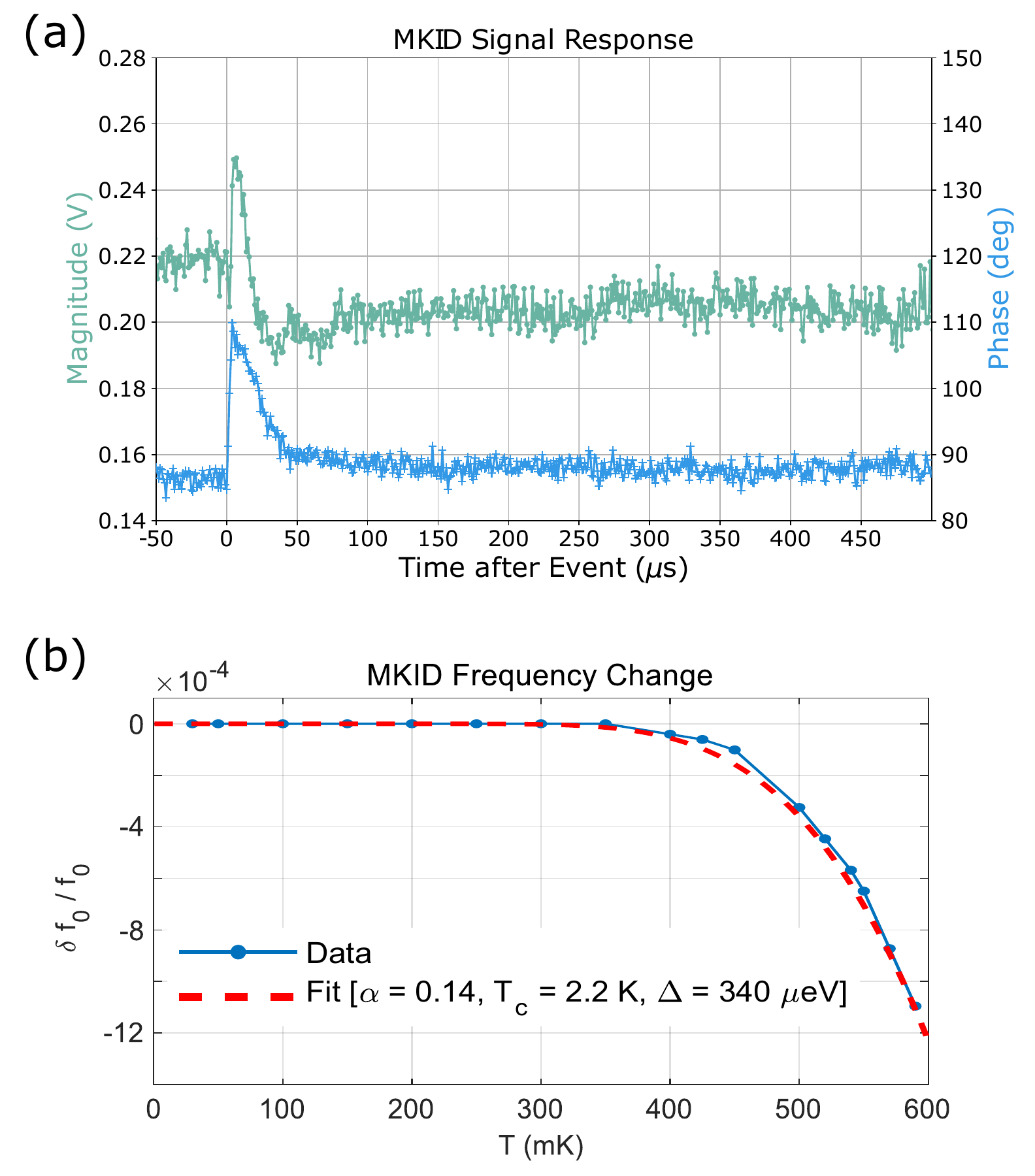}
	\caption{\label{fig:Figure2} MKID response curves. (a) The MKID magnitude change (top, green) and phase change (bottom, blue) following an ionizing radiation event. The S21 is an average over 1 $\mu$s, and is used as a the trigger for an on-chip ionizing radiation event. The fast recovery characteristic time is 35 $\mu$s, which sets the timescale for resetting the sensor, with a slower secondary recovery time of a couple milliseconds. A matched filter is used to provide the timestamp of an ionizing radiation event used as the reference time for the qubit measurements in Fig.~\ref{fig:Figure3}. (b) MKID frequency response to temperature. The calibration curve fit allows for estimates of the kinetic induction fraction ($\alpha$~=~0.14), the critical temperature ($T_c$~=~2.2~K), and superconducting gap ($\Delta$ = 340 $\mu$eV). From this data, we estimate a frequency shift of 0.67 Hz/QP, suggesting the MKID needs 25 eV of total energy deposited in the grAl film to register a sufficient signal.}
\end{figure}

To detect ionizing radiation, we use MKIDs, which are resonant structures fabricated with high-kinetic inductance films \cite{gao_thesis}. Broken Cooper pairs in the MKID film cause an increase in the inductance and a corresponding decrease in the resonant frequency. We use 30-nm-thick films of granular aluminum (grAl) for the MKID superconducting material, and we follow the design concept from Ref.~\cite{Valenti_2019}. We fabricate four MKIDs in each quadrant of a $6.7 \times 6.7$\, mm$^2$ reticle on a 675-$\mu$m-thick silicon chip (Fig.~\ref{fig:Figure1}), and all MKIDs are capacitively coupled to a single feedline, with an external quality factor of 50k. We estimate the internal quality factor of the MKIDs at 30k, which results in a measured linewidth of 250 kHz (FWMH), allowing for microsecond signal response (Appendix~\ref{app:time}). To read out the MKIDs, a continuous waveform (CW) is applied to the feedline and averaged every $\mu$s to provide a quasi-continuous S21 trace. From thermal characterization of the MKIDs (Fig.~\ref{fig:Figure2}b), we estimate they require at least 40 keV of deposited energy, or $7.5\times 10^4$ QPs, to shift the MKID resonance frequency by a quarter of the linewidth. The MKIDs exhibit a fast recovery characteristic time of 35 $\mu$s, which is well-fit by a decaying exponential. While a slower recovery time is also observed, the fast dynamics are sufficient to allow the MKID to retrigger after 20 $\mu$s following an event, an example trace is shown in  Fig.~\ref{fig:Figure2}a.

The average impact radius following an ionizing radiation event can be estimated through correlated event detections of the MKIDs. We used three MKIDs (red highlight in Fig.~\ref{fig:Figure1}a) to measure simultaneous detection events. The conditional probabilities are shown in  Fig.~\ref{fig:Figure1}b. The measured conditional probabilities are consistent with energy from ionizing radiation propagating throughout the entire chip, and MKID detection efficiencies of 0.90, 0.89, and 0.50 for MKID 1b, 1d, and 4c, respectively. This sets a lower-bound radius at 2.3~mm, limited by the largest distance between the measured MKIDs. The relative efficiency between the MKID 1b (or 1d) and 4b can be explained by the differences in the MKID surface areas, where MKIDs 1b and 1d have a surface area of $2.8\times 10^4~\mu m^2$, and MKID 4b has a surface area of $2.0\times 10^4~\mu m^2$, which would predict a relative efficiency of 0.73 before accounting for the higher rate of lower-energy ionizing radiation events.

Eight qubits surround each set of MKIDs, as shown in Fig. \ref{fig:Figure1}a. The ground plane is defined with niobium films capped with a thin film of aluminum, the qubit capacitors are entirely niobium, and qubit junctions are made with Al/AlOx/Al using a standard Dolan bridge process. The transmons are a single-junction design with $E_j/E_c = 350$ and $f_q=11$ GHz. To assess the impact on the qubits, we prepare the qubit in the desired state ($|0\rangle$ or $|1\rangle$), followed by a 1~$\mu$s idle before measuring the final state. The readout tone that discriminates the qubit state is averaged for 1 $\mu$s, similar to the MKID. The transmon is then thermally reset for 50 $\mu$s before repeating the sequence.

\begin{figure}[t]
	\includegraphics[width=0.5\textwidth]{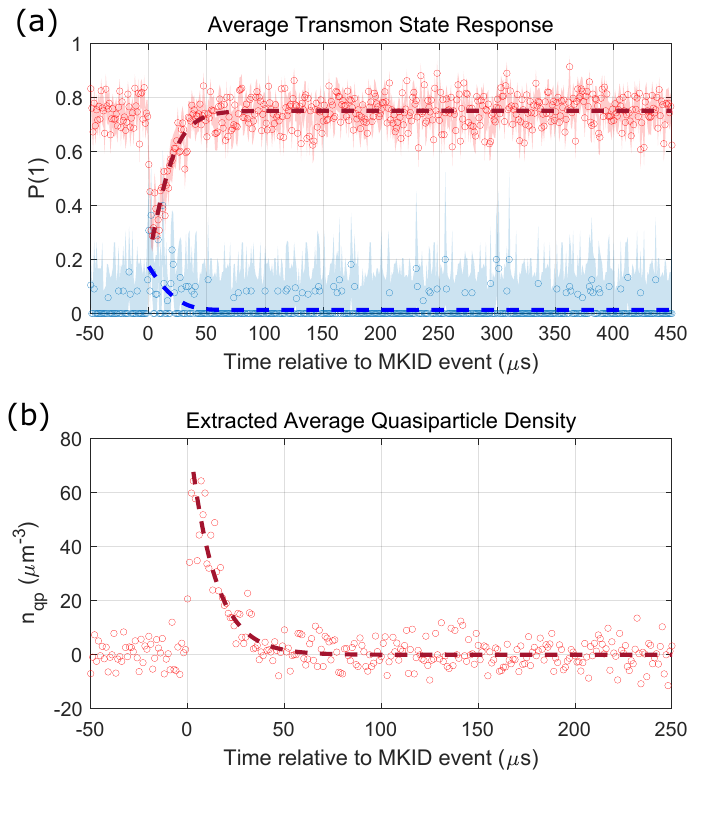}
	\caption{\label{fig:Figure3}  (a) Response of the transmon $|1\rangle$-state (red) and $|0\rangle$-state (blue) to an ionizing radiation event. By continuously streaming a measurement of an MKID positioned near a transmon, and averaging over 3731 (606) events during 90 (17) hours of data collection, the averaged transmon response for enhanced decay (excitations) can be determined. Our recovery time for the enhanced qubit decay (excitation) following an event has a characteristic time of $13\pm1\ \mu$s ($8.3^{+1.7}_{-1.2}\ \mu$s) in our Nb-based devices. (b) The extracted average quasiparticle density at the junction from $P(1)$. We infer a peak quaisparticle density, ($n_{\rm qp}(\delta\text{t=0})$), is $85^{+9}_{-8} / \mu m^{-3}$, averaging over all detected events.}
\end{figure}

The qubit sequence and MKID measurements are run simultaneously. The MKID stream is processed live to search for candidate ionizing radiation events (Appendix~\ref{app:exp}). If a signal is registered, the entire buffer of both the MKID and qubit streams are saved for additional analysis. For each detected event, all qubit measurements performed near the identified event are tallied at each $\delta t$ relative to the MKID event. In this way, the qubit excited state probability can be estimated for each $\delta t$ relative to an ionizing radiation event with $\mu$s resolution. The resultant data are shown in Fig.~\ref{fig:Figure3}.

\section{\label{sec:simulation}Ionizing Radiation Simulation}

The expected cosmic ray interaction rate and deposited energy in the silicon are predicted using the GEANT4 \cite{GEANT4_03,gEANT4_06,GEANT4_16} toolkit to model the laboratory and fridge environment down to the silicon chip and transport cosmic rays at sea level as generated by the Cosmic-Ray Shower Library (CRY) software \cite{osti_902609,cry_tech}. The simulation includes all available particles: muons, protons, neutrons, electrons, positrons, photons, and pions. The CRY simulation generated cosmic rays assuming the maximum flux during the solar cycle over a 20 $\times$ 20 $\text{m}^2$ area.

The cosmic rays become inputs to the GEANT4 model entering at an approximation of the material above the fridge setup, and propagate into a 10 $\times$ 10 $\text{m}^2$ box, 6\,m tall and complete with material for ceiling, walls, and floor to account for particles incident at large zenith angles. The dilution refrigerator was modeled including the shielding for each stage, the helium-3 inlet, dilution unit, cold finger surrounded by mu-metal shielding and containing the aluminum chip package which holds the $6.7 \times 6.7$\, mm$^2 \times$ 675-$\mu$m silicon chip.

The particles deposit energy ranging from 0.1\,keV to 12\,MeV with a mean (median) of 340\,keV (260 keV), and a standard deviation of the distribution of 310\,keV. For comparison the average cosmic ray muon deposits 460\,keV. For the threshold energy of our MKID (40~keV), the expected average energy deposit across all particles is 350~keV, and the predicted MKID trigger rate is one event per 131~s, neglecting environmental gammas.

\section{\label{sec:recovery}Recovery Dynamics}

To assess the qubit relaxation, we collected data for 90 hours, resulting in 3731 total ionizing radiation events. We observed a sudden decrease in the measured excited state population after an ionizing radiation event. The recovery to baseline following an event fits well to a QP trapping model ($\dot{n}_{qp}=-s n_{qp}$), with a trapping characteristic time $1/s = 13\pm1\ \mu$s. From the $T_1$ at the peak of the ionizing radiation event, we estimate a quasiparticle density of $n_{qp}= 85^{+9}_{-8} / \mu m^{-3}$ averaged over all detected events. Taken with our average energy deposit of detected events (350 keV), we estimate a quasiparticle density at the junction per deposited energy of 240\,$/\mu m^3/$MeV.

To probe the qubit excitation dynamics, we allow the qubit to thermally relax to the ground state before measurement. We observed 606 ionizing radiation events over 17 hours. Counter to other literature results \cite{McEwen_2021}, the measurements show an increased excited state probability following a radiation event. We find the characteristic recovery time of the enhanced excitations to be $8.3^{+1.7}_{-1.2}\ \mu$s. The observed excitations are likely due to the proximity of Nb traces to our junction, where we have approximately 2.5 $\mu$m of Al between the junction barrier and the Nb trace, which likely does not give the QPs sufficient distance to downconvert from the Nb band edge to the Al band edge before arriving to the junction (Appendix~\ref{app:eect}).

It is interesting to compare our measurements of recovery time with similar published results \cite{Wilen_2021,Yelton_2024,McEwen_2021,harrington2024synchronousdetectioncosmicrays}, where reported recovery times range from $25~\mu\text{s}$ to 5~ms. Here, we extend this range slightly lower to $13~ \mu\text{s}$. There is a broad range of related time scales caused by various theoretical mechanisms that include phonon and quasiparticle transport, phonon lifetimes and quasiparticle lifetimes. Variations in recovery time might result from differences in chip geometry, junction geometry, materials, and microstructure, but also differences in quasiparticle and phonon energies. Analysis of phonon transport times and phonon lifetimes in Nb and Al suggests that, after the first microsecond, most excess energy is stored as excess QPs with only a small fraction stored as phonons (Appendix~\ref{app:plis}). Consequently, in all cases, quasiparticle transport or quasiparticle lifetimes must be considered.

We now turn to a more focused comparison between our measurements and those from Yelton et al.~\cite{Yelton_2024}, where they observed very similar recovery times even though the injected phonons were at different energies than would be expected from ionizing radiation. In Ref.~\cite{Yelton_2024}, QPs were injected with Al/AlOx/Al junctions, and thus at the aluminum band edge. This injection would result in 2$\Delta_{\text{Al}}$ phonons in the silicon with insufficient energy to break Cooper pairs in the Nb qubit capacitors and ground planes, so the phonons can only interact with the qubit through breaking Cooper pairs in the junction lead and then generating QPs at the Al band edge. This selective Cooper-pair breaking was confirmed by the absence of qubit excitation errors, $|0\rangle \rightarrow |1\rangle$.  In our system, injected phonons have sufficient energy to break Nb Cooper pairs; therefore the energy can undergo Nb diffusion and trapping. Despite this difference, we observe similar recovery dynamics for $P(1)$, where both data sets fit well to a trapping model with similar trapping times, $13~\mu\text{s}$ here compared to $25~\mu\text{s}$ in Ref.~\cite{Yelton_2024}. As the QPs in our device can form in both Nb and Al, whereas the QPs in Ref.~\cite{Yelton_2024} form only in the Al, the closeness of the quasiparticle dynamics suggests that $P(1)$ recovery is determined by quasiparticle lifetimes in the Al leads, with the dynamics most likely driven by trapping.

The presence of Nb, however, appears to play an essential role in hastening $T_1$ recovery. The superconducting proximity effect may contribute to trapping in a way that explains why transmons with Nb and Al have $T_1$ recovery times on the low end of the range, tens of $\mu$s, not ms. Junctions with much longer Al leads as reported in Ref.~\cite{Wilen_2021} (compared to the junctions here and in Ref.~\cite{Yelton_2024}) will have greatly reduced impact of the superconducting proximity effect near the junction, and the $T_1$ recovery time is much longer ($130~\mu\text{s}$). The longest observed $T_1$ recovery times occur in Refs.~\cite{McEwen_2021,harrington2024synchronousdetectioncosmicrays} ($25~\text{ms}$ and $6~\text{ms}$, respectively) where there is no proximity effect; the transmons are entirely Al. It is important to note that there are multiple simultaneous transport and reaction processes that can make quantitative prediction difficult, but the observed impact of the presence of both Nb and Al in transmons is quite strong.

\section{\label{sec:tls}TLS Dynamics}

\begin{figure}[b]
	\includegraphics[width=0.40\textwidth]{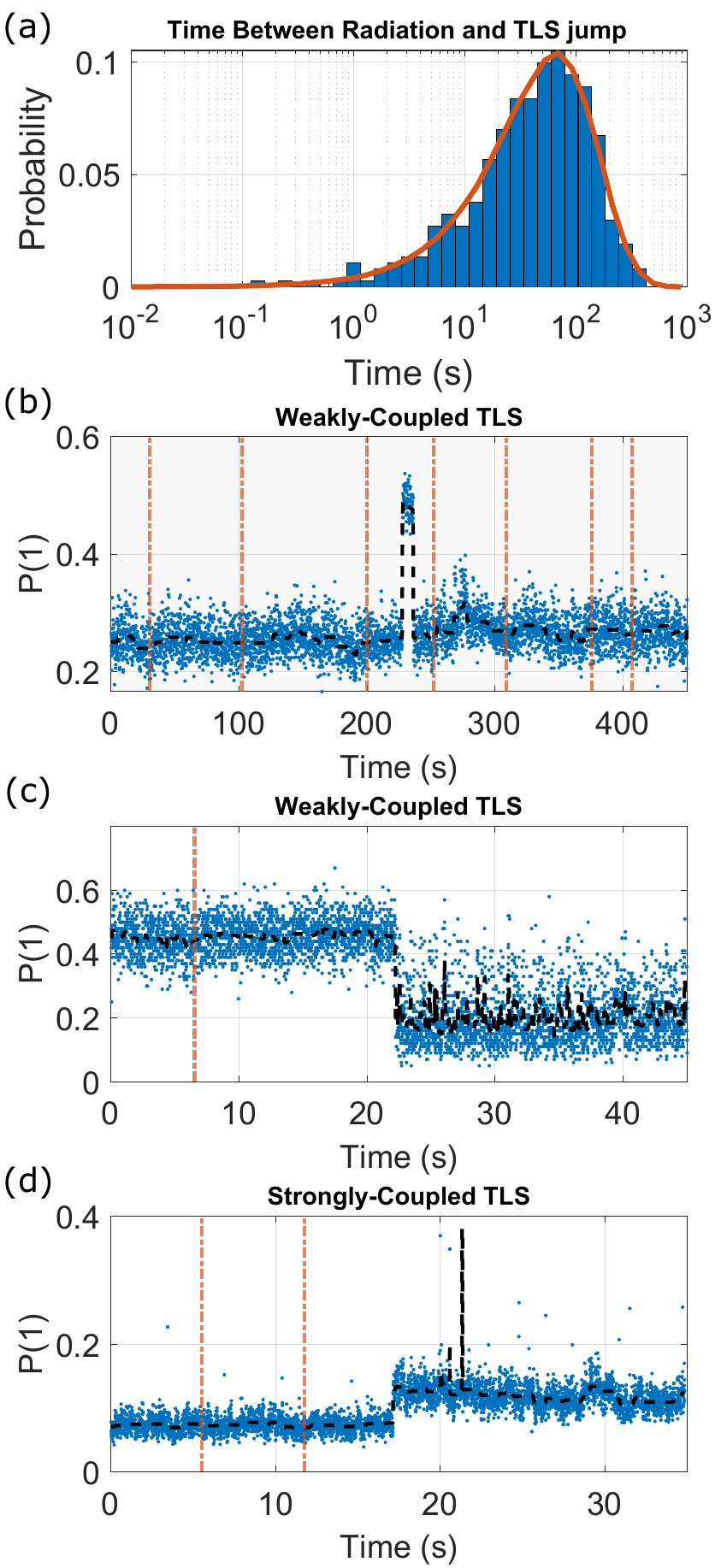}
	\caption{\label{fig:Figure4} TLS dynamics following ionizing radiation events. (a) Time difference of the closest ionizing radiation event that precedes a TLS scrambling event. The blue bars are the observed time difference. The orange line is the expected curve for uncorrelated data. None of the 18,454 ionizing radiation events occurred within 100 ms of any of the 371 TLS scrambling events. (b-d) Example time dynamics of a weakly-coupled (b-c) and strongly-coupled (d) TLS. The orange vertical dashed lines are events detected by the MKID. The blue dots are the data averaged over 100 ms (b,d) and 10 ms (c). The black curve is a further averaged window to provide a visual guide.}
\end{figure}

To determine the correlation between TLS dynamics and ionizing radiation, we identify two unique sets of defects. The first set of defects are identified with MHz-wide peaks in the qubit $\Gamma_1$ spectrum, with fits to the qubit relaxation rates providing estimates of coupling strengths near 100 kHz \cite{Barends_2013}. These defects are likely located on the edges of the Nb capacitor pads, and likely originate in the Nb-Si or the Nb-Air interface. The second set was identified by exploring large frequency windows to find Rabi oscillations in the $T_1$ spectroscopy. These defects have coupling strengths near 100 MHz and originate inside the junction oxide. The qubit is biased to be on resonance with the TLS and $P(1)$ is monitored as done previously --- we prepare the qubit in the $|1\rangle$ state following by a short idle and readout and then allow for thermal reset. We again simultaneously continuously measure the MKIDs near the qubit, with each defect being monitored for up to 24 h to create a time trace of $P(1)$. Large and sudden changes in $P(1)$ are labeled as TLS scrambling events, which are compared to ionizing radiation events detected from the MKID. A selection of representative curves is shown in Fig.~\ref{fig:Figure4}b-d.

We observed 371 total TLS scrambling events and 18,454 detected radiation events across 430 hours of measurements. In contrast to previous reported results, we do not find any correlations between our MKID events (red dashed lines) and TLS scrambling events. Across all events, the closest preceding radiation event was 126 ms before the TLS event, after subtracting timing uncertainties from data averaging. A key difference between our measurements and Thorbeck et al. \cite{Thorbeck_2023} is the energy threshold required to register a signal. Our MKIDs require approximately 25 eV of energy in the form of broken Cooper pairs in MKID to register a signal. Ionizing radiation will routinely deposit the necessary 40 keV of energy on the chip to achieve the energy threshold. Charge-sensitive transmons, however, are sensitive to single-electron trapping in the charge-sensing volume of the qubit. Such charge-trapping events could be triggered by events that release substantially less energy than the amount ionizing radiation would deposit. For example, strain relaxation could produce bursts of energy insufficient to trigger our MKIDs \cite{Anthony_Petersen_2024}, but may be observed by a charge-sensitive transmon. This could further explain residual radiation-like events observed in an underground facility \cite{bratrud2024measurementcorrelatedchargenoise}.

Measurements have also shown that the rate of quasiparticle bursts decreasing by an order of magnitude over time \cite{Mannila_2021}, indicating that events are occurring in the cryogenic space, which produce radiation-like signals. Anomalous behavior has been seen in other cryogenic experiments, where an abundance of low-energy background (LEE) with energy scales up to 100 eV has been observed. LEE has been shown to correlate with device strain \cite{Mannila_2021}, and a similar decreases of rate over time has been observed, with Ref.~\cite{Anthony_Petersen_2024} seeing a 0.5 dB/day decrease in rate of LEE, compared to Ref.~\cite{Mannila_2021} observing a 0.1 dB/day decrease of the rate of QP bursts. While there is no evidence of correlations between LEE and charge jumps, it is worth noting the ubiquity of observed complex physics in cryogenic experiments that are unrelated to ionizing radiation.

\section{\label{sec:conclusions}Conclusion}

We have integrated an MKID array directly on a device with superconducting qubits, which allows for measuring the impact of ionizing radiation events on qubit performance with microsecond timing resolution. Using the MKID array, we observed that radiation events appeared to be chip-wide on the $6.7\times 6.7~\text{mm}^2$ die. We simultaneously measured ionizing radiation events and TLS dynamics, monitoring both strongly- and weakly-coupled defects. We find no correlation between ionizing radiation events and discreet TLS scrambling events. This result suggests that the observed charge jumps correlated with TLS scrambling events may be due to different underlying physics, such as strain relaxation. Our observed discrepancy is further reinforced by previous measurements in similar devices \cite{bratrud2024measurementcorrelatedchargenoise,Mannila_2021}, which highlight other complex, radiation-like, phenomena occurring in devices at cryogenic temperatures.

We also characterized the enhanced decay and excess excitations in the presence of an ionizing radiation event. From our averaged qubit relaxation measurements, we observe a quasiparticle density in our qubit junctions leads of $n_{qp}= 85^{+9}_{-8} / \mu m^{-3}$ averaging over all ionizing radiation events. From our MKID temperature characterization, we estimate that ionizing radiation thus causes a quasiparticle density spike of 240\,$/\mu m^3/$MeV, with a characteristic recovery time of $13\pm1\ \mu s$ ($8.3^{+1.7}_{-1.2}\ \mu$s) for recovery of the enhanced decay (excess excitations). From a comparison of our results with literature (\cite{Yelton_2024,Wilen_2021,harrington2024synchronousdetectioncosmicrays,McEwen_2021}), we posit that the proximity of niobium to the junction drives the recovery dynamics from quasiparticle bursts.

These results showcase the complexity of superconducting circuits and their cryogenic environment. Further study of test platforms with integrated radiation detection \cite{Castelli_2025} would allow a better understanding of the underlying dynamics. In particular, the dynamics of TLS in the presence of radiation, and the dependence of the recovery timescales of the quasiparticle bursts on the material stacks. Our device demonstrates the ease of integrating radiation detection on the same substrate that carries the qubits, providing a simple test platform for characterizing radiation impacts without requiring adding any additional materials to the standard process flow.

\begin{acknowledgments}
The authors thank Thomas Chamberlin, Michael Chilcote, Dylan Cowger, Nikolaus Hartman, Bob Hinkey, Moe Khalil, Chan Kim, Alex Krick, Christopher Langlett, Edward Leonard, James Medford, JT Mlack, Roy Murray, Kelin Kelcourse-Oquendo, Paige Quarterman, Andrew Schwarzkopf, Ryan Stein, Joel Strand, Ian Thompson, Jon Vannucci, Owen Vils, Michael Wayne, and Seth Whitsitt for support.

This work was supported in part by the United States Government (USG).
\end{acknowledgments}

\clearpage
\newpage
\setcounter{figure}{0}
\renewcommand{\thefigure}{S\arabic{figure}}
\renewcommand{\theequation}{S\arabic{equation}}
\renewcommand{\thetable}{S\arabic{table}}

\appendix

\section{Experimental Setup}
\label{app:exp}

The experiments are conducted in a dry dilution refrigerator operated at around 30 mK. The sample is enclosed in a mu-metal can for magnetic shielding at the mixing chamber stage. All rf drive lines (MKID, readout, etc.) are attenuated by 20~dB at the 3-K, 800-mK (still), and 30-mK (mixing chamber) stages, before passing through Eccosorb$^{\textregistered}$ low pass filters (Eccosorb is a registered trademark of Laird Technologies). The return signals are amplified by a traveling-wave parametric amplifiers (TWPA) at the 30-mK stage, followed by a high electron mobility transistor (HEMT) amplifier at the 3-K stage. Room-temperature amplifiers are used for another 36~dB of gain before passing through an IQ mixer to downconvert to 10 MHz. Finally, the signals are digitized on an AlazarTech$^{\textregistered}$ ATS9416 card (AlazarTech ATS is a registered trademark of Alazar Technologies Inc.).

To search for MKID events, two event detecting schemes are used: live event detection to minimize the data storage, and an off-line event detection for better fidelity detection and timing resolution. The live event detector uses the raw data to look for a non-symmetric deviation in IQ space, which is a characteristic feature of the event. When the center of mass of outer data is located about 3 standard deviations away from the center, it is marked as a possible event and the data are saved. The off-line event detector is applied to the previously saved data to filter out false positive events with more stringent limits. The MKID amplitude trace is processed through low pass and high pass filters followed by matched filter with a decaying exponential template. If the processed data has values outside of a threshold (8 standard deviations), it is marked as an event for final event counting.

\section{Measurement Response Time}

\begin{figure}[h]
	\includegraphics[scale=0.28]{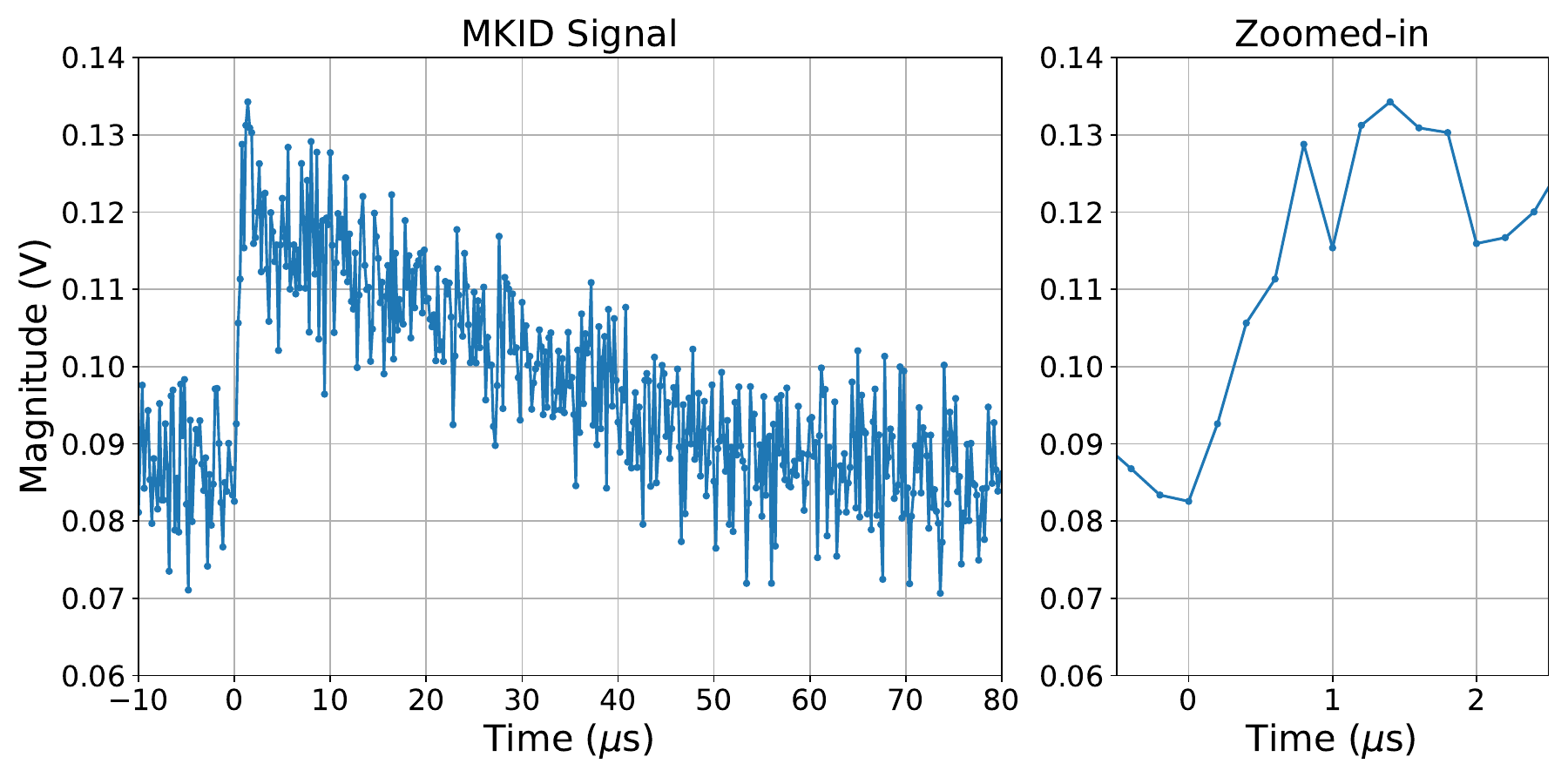}
	\caption{\label{fig:Figure5} MKID signal vs time relative to event. The MKID $|\text{S21}|$ near an event with 200-ns averaged time-bins, with an 80-$\mu$s-long trace (left) and a zoomed-in 3-$\mu$s-long trace (right). The MKID response shows a rising edge of 600--700 ns, matching the expectation from the measured resonance linewidth (250 kHz).}
\end{figure}

The digitizer used to collect the data has a sampling rate of 100 MS/s. The data presented in the main text is collected and processed live with a readout integration time of 1 $\mu$s. Fig.~\ref{fig:Figure5} shows the MKID response near an event with a 200 ns average. The rising edge after the event (at 0 $\mu$s) exhibits a slope indicative of a  600--700 ns response time of the MKID, consistent with the measured 250 kHz linewidth.

\label{app:time}

\section{Excess Excitation Characteristic Time}
\label{app:eect}
The lifetime of quasiparticle of energy $E = \Delta_\text{Nb} = 1.40~\text{meV}$ in Al is found from Eq. 7 in Ref.~\cite{Kaplan_1976}. At a temperature of $20~\text{mK}$, the energetic quasiparticle lifetime is $0.6~\text{ns}$, much smaller than the measured value from prep zero measurements, $8.3~\mu\text{s}$. Therefore, the quasiparticles that cause $|0\rangle \rightarrow |1\rangle$ transitions are likely much lower energy than $\Delta_\text{Nb}$. At an energy level of $\Delta_\text{Al} + (53~\mu\text{eV})$, the quasiparticle lifetime matches the measured prep zero recovery time. From this crude analysis, it is expected that the quasiparticles causing the excitations follow a broad distribution of quasiparticle energies arising via a multistep thermalization process.

\section{Phonon Lifetime in Si}
\label{app:plis}

The chips measured here, as well as those discussed in Refs.~\cite{Wilen_2021,Yelton_2024,McEwen_2021,harrington2024synchronousdetectioncosmicrays}, have a high density of superconducting ground planes. The mean lifetime of phonons originating in Si substrates is determined by the time for a phonon to reach the ground plane at the top of each chip, the probability of transmission into the metal ground plane and the lifetime of a phonon in the metal ground plane. For a rough estimate, the isotropic approximation is suitable, and phonons are assumed to be uniformly distributed through the height of the substrate. The average upward phonon velocity is well-known and found by integrating over the phonon direction solid angle to get $v/4$, where $v$ is the average phonon velocity~\cite{Swartz_1989}. The mean time for phonons in a substrate of height $h$ to impinge on the substrate metal interface is $4h/v$. The mean number of trips is $1/P_\text{trans}$, the transmission probability. The mean phonon lifetime is found by adding the Cooper-pair breaking lifetime \cite{Kaplan_1976} to the phonon travel time.
\begin{equation}
\tau^\text{ph} = \frac{4 h}{v P_\text{trans}} + \tau_\text{b} 
\end{equation}

We perform these calculations for both Al and Nb ground planes. The Cooper-pair breaking lifetime at low temperatures is bounded from above by the characteristic phonon lifetime $\tau^\text{ph}_0$ \cite{Kaplan_1976}. For Al, $\tau_\text{b}<0.242~\text{ns}$; for Nb, $\tau_\text{b}<4.17~\text{ps}$, negligible for calculating $\tau^\text{ph}$. The phonon transmission probability can be estimated using the diffuse mismatch model (DMM) \cite{Swartz_1989}. For Si, the longitudinal and transverse phonon velocities are $v_\text{L} = 8970~\text{m/s} $ and $v_\text{T}=5332~\text{m/s}$ \cite{Swartz_1989}. The mean phonon velocity is $v = (v_\text{L}^{-2} + 2 v_\text{T}^{-2})/ (v_\text{L}^{-3} + 2 v_\text{T}^{-3}) = 6408~\text{m/s}$. For Al, the phonon velocities are $v_\text{L} = 6808~\text{m/s} $ and $v_\text{T}=3251~\text{m/s}$ \cite{Yelton_2024}. For Nb, $v_\text{L} = 5139~\text{m/s}$ and $v_\text{T}=2168~\text{m/s}$ \cite{Yelton_2024}. Using the DMM, the transmission probabilities from material 1 to material 2 are $P_\text{trans}=(v_\text{2,L}^{-2}+2 v_\text{2,T}^{-2})/[(v_\text{1,L}^{-2}+2 v_\text{1,T}^{-2})+(v_\text{2,L}^{-2}+2 v_\text{2,T}^{-2})]$ \cite{Swartz_1989}. Phonon transmission from Si into Al has a probability $P_\text{trans}=0.78$. For Nb ground planes, $P_\text{trans}=0.91$. Using a characteristic Si substrate height of $h \sim 500~\mu\text{m}$, the phonon lifetime estimates are $\tau^\text{ph}\approx 400~\text{ns}$ for Al and $\tau^\text{ph}\approx 340~\text{ns}$ for Nb, lifetimes that are far smaller than the range of interest, $8.3~\mu\text{s} ~\text{to}~ 25~\text{ms}$.

\bibliographystyle{unsrt}
\bibliography{references}

\end{document}